\documentclass{aa}    

\usepackage{graphicx,url,twoopt,natbib}
\usepackage[varg]{txfonts}           
\usepackage{hyperref}                
\usepackage{pdfcomment,acronym}      
\hypersetup{
  urlcolor=blue,     
  linkcolor=red,     
}

\makeatletter
\newcommand{\bibnote}[2]{\global\@namedef{#1note}{#2}}
\newcommand{\biblink}[2]{\global\@namedef{#1link}{#2}}
\makeatother

\makeatletter
 \newcommandtwoopt{\citeads}[3][][]{%
   \nonstopmode
   \href{http://adsabs.harvard.edu/abs/#3}%
        {\def\hyper@linkstart##1##2{}%
         \let\hyper@linkend\@empty\citealp[#1][#2]{#3}}
   \biblink{#3}{\href{http://adsabs.harvard.edu/abs/#3}{ADS}}%
   \errorstopmode}            
 \newcommandtwoopt{\citepads}[3][][]{%
   \nonstopmode
   \href{http://adsabs.harvard.edu/abs/#3}%
        {\def\hyper@linkstart##1##2{}%
         \let\hyper@linkend\@empty\citep[#1][#2]{#3}}
   \biblink{#3}{\href{http://adsabs.harvard.edu/abs/#3}{ADS}}%
   \errorstopmode}            
 \newcommandtwoopt{\citetads}[3][][]{%
   \nonstopmode
   \href{http://adsabs.harvard.edu/abs/#3}%
        {\def\hyper@linkstart##1##2{}%
         \let\hyper@linkend\@empty\citet[#1][#2]{#3}}
   \biblink{#3}{\href{http://adsabs.harvard.edu/abs/#3}{ADS}}%
   \errorstopmode}            
 \newcommandtwoopt{\citeyearads}[3][][]{%
   \nonstopmode
   \href{http://adsabs.harvard.edu/abs/#3}%
        {\def\hyper@linkstart##1##2{}%
         \let\hyper@linkend\@empty\citeyear[#1][#2]{#3}}
   \biblink{#3}{\href{http://adsabs.harvard.edu/abs/#3}{ADS}}%
   \errorstopmode}            
\makeatother

\def\linkadspage#1#2#3{\href{http://adsabs.harvard.edu/cgi-bin/nph-data_query?bibcode=#1\&link_type=ARTICLE\&d_key=AST\#page=#2}{#3}}

\begin{document}  

\title{The discrepancy between dynamical and theoretical mass in the triplet-system 2MASS J10364483+1521394}

\author{Per Calissendorff \inst{1}
\and Markus Janson\inst{1}
\and Rainer K{\"o}hler\inst{2, 3}
\and Stephen Durkan\inst{4}
\and Stefan Hippler\inst{5}
\and Xiaolin Dai\inst{5}
\and Wolfgang Brandner\inst{5}
\and Joshua Schlieder\inst{6}
\and Thomas Henning\inst{5}
}
				
\institute{Department of Astronomy, Stockholm University, Stockholm, Sweden \
\and Institut f\"{u}r Astro- und Teilchenphysik, Universit\"{a}t Innsbruck, Technikerstr. 25/8, 6020 Innsbruck, Austria\
\and Department of Astrophysics, University of Vienna, Vienna, Austria\
\and Astrophysics Research Centre, Queens University, Belfast, Belfast, Northern Ireland, UK\
\and Max Planck Institute for Astronomy, Heidelberg, Germany
\and NASA Goddard Space Flight Center, Greenbelt, Maryland, USA
}				


\abstract{We combine new Lucky Imaging astrometry from NTT/AstraLux Sur with already published astrometry from the AstraLux Large M-dwarf Multiplicity Survey to compute orbital elements and individual masses of the 2MASS J10364483+1521394 triple system belonging to the Ursa-Major moving group. The system consists of one primary low-mass M-dwarf orbited by two less massive companions, for which we determine a combined dynamical mass of $M_{\rm{B}+\rm{C}}= 0.48 \pm 0.14\ M_\odot$. We show from the companions relative motions that they are of equal mass (with a mass ratio of $1.00 \pm 0.03$), thus $0.24 \pm 0.07\ M_\odot$ individually, with a separation of $3.2 \pm 0.3\ $AU and conclude that these masses are significantly higher ($30\%$) than what is predicted by theoretical stellar evolutionary models. The biggest uncertainty remains the distance to the system, here adopted as $20.1 \pm 2.0$ pc based on trigonometric parallax, whose ambiguity has a major impact on the result.  With the new observational data we are able to conclude that the orbital period of the BC pair is $8.41^{+0.04}_{-0.02}\ $years. 
}


\titlerunning{Mass discrepancy 2M1036}
\authorrunning{Calissendorff et al.}

\maketitle

\section{Introduction}     \label{sec:introduction}

The importance of stellar multiplicity studies have long been recognised, and given that multiplicity is a common natural process of star formation, we have ample opportunities to further improve such studies. By monitoring the motions of multiple systems we can empirically constrain their dynamical masses, which become invaluable resources for theoretical models of stellar formation and evolution \citep{Goodwin, Bate}. There has lately been a renewed cause to study the low-mass regime of multiplicity as the characterisation of the nearby M-dwarf population is becoming more complete \citep{Riaz, Reid, Riedel}. \citet{Duchene} estimate the multiplicity fraction of low-mass stars ($0.1 - 0.5\ M_\odot$) to be $26 \pm 3 \%$, and that the mass ratio between primary and secondary is too intricate to be described by a single power-law distribution, and depends on both primary mass and binary separation.

\begin{figure}[hbtp]
  \centering
  \includegraphics[width=\columnwidth]{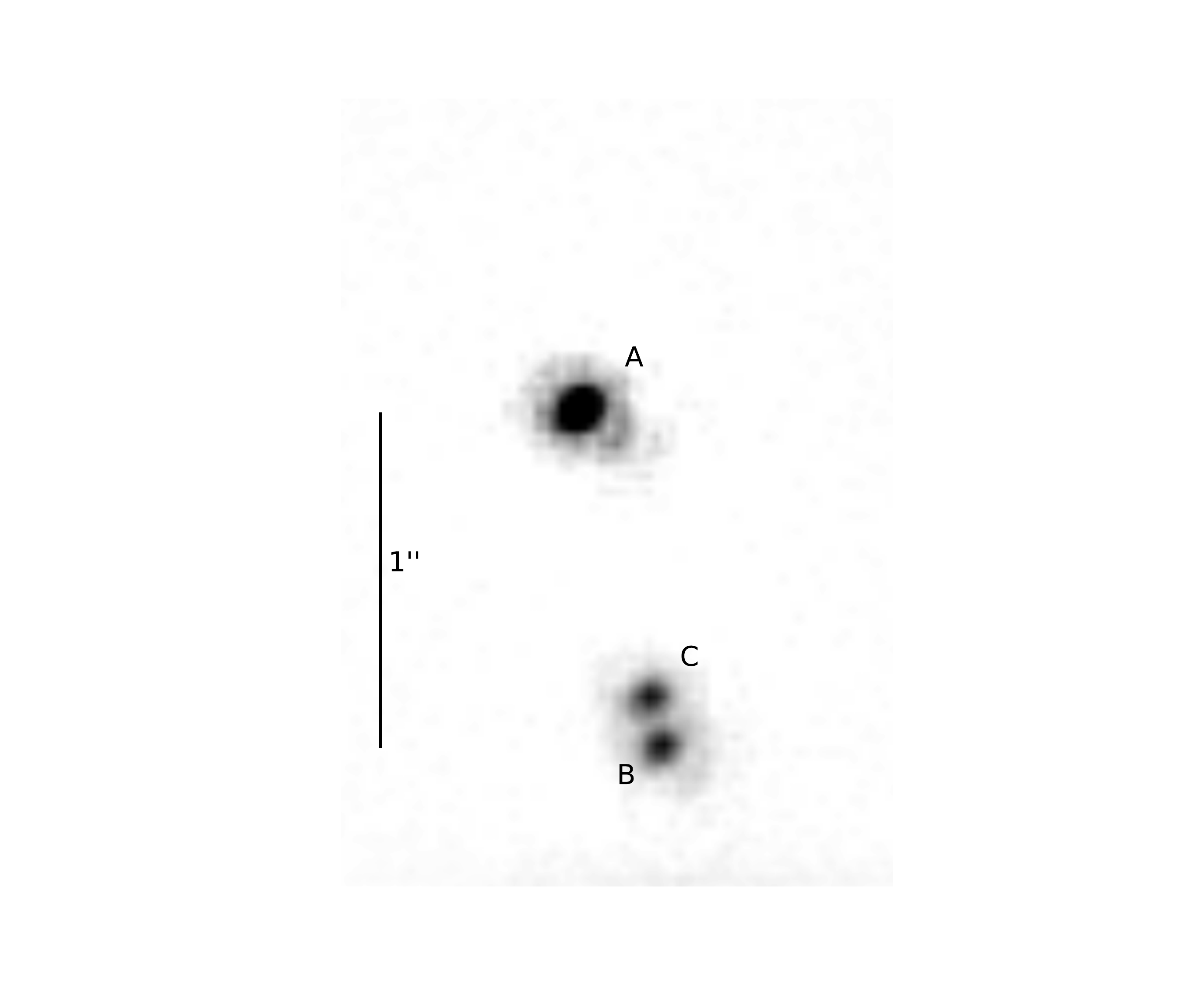}
  \caption[]{\label{fig:J1036} %
    Reduced $z'$-filter image of 2MASS J10364483+1521394 obtained with AstraLux Norte at Calar Alto, Spain, in November 2008. The separation between A and B is roughly 1'', and the separation between B and C is $\sim150\ $mas (cf. Table~\ref{tab:astrometry}). The orientation in the image is North up and East to the left.
  }
\end{figure}

The AstraLux Multiplicity Survey \citep{Janson 2012, Janson 2014} has monitored several low-mass systems, one of which is the triplet system 2MASS J10364483+1521394 (subsequently J1036). The system consists of one low-mass primary A with two companions B and C orbiting A at a distance of $\approx 1''$, separated by roughly $0.16'' - 0.18''$ themselves (see Figure~\ref{fig:J1036}). There are hints that the system is associated with the Ursa-Major Moving Group (UMa MG) \citep{Klutsch}, but distance measurements to the system remain ambiguous, ranging from 7 to 20 pc depending on the technique employed \citep{Riaz, Daemgen, Shkolnik}. Here, we adopt the  trigonometric parallax distance estimate from  \citet{Shkolnik} as $d = 20.1 \pm 2.0\ $pc, as it should be the more reliable estimate compared to previous unresolved measurements.

The AstraLux campaign monitored the orbit of J1036ABC for almost a decade using the Lucky Imaging technique. By relying on serendipitous short exposures, the observations become nearly diffraction limited rather than limited by the seeing. Probing the orbital elements of the system yields an opportunity to estimate the dynamical mass of each individual component of the system. During the orbital monitoring of J1036, the companion components B and C have completed a full orbital cycle, thus making it possible to constrain the orbital period very well. The orbit of BC around A on the other hand is not yet fully known, but is likely to be on the order of a couple of centuries.

New relative astrometric measurements of J1036 by the AstraLux Multiplicity Survey are presented in this paper, which we combine with older measurements from both the survey and other literature in order to derive estimates for the orbital parameters of the system. We also discuss how the results compare theoretical models and note that the system is not well represented by the current stellar evolutionary track models.

\section{Observations and data reduction}    \label{sec:observations}
Observations collected for this work span roughly a decade, primarily based on exposures with the AstraLux\footnote{
\url{http://www.mpia.de/ASTRALUX/} }
 Norte camera \citep{Hormuth} on the 2.2m telescope at Calar Alto in Spain, along with the AstraLux Sur camera \citep{Hippler} on the ESO/NTT 3.5m telescope at La Silla in Chile. We obtain one of the epochs in this study from \citet{Daemgen}\footnote{The position angle is shifted exactly $360\ ^{\circ}$ compared to the literature value \citep[see \linkadspage{2007ApJ...654..558D}{5}{Table 2} in][]{Daemgen}.}, where the Gemini North$/$Altair natural guide star adaptive optics was used. A more detailed description of the individual astrometry measurements is presented in Table~\ref{tab:astrometry}.

AstraLux multiplicity observations employ the same procedures as described by \citet{Daemgen 2009, Bergfors 2013} where a total integration time of 300\ s is obtained by adding up short individual readouts, typically of 15-30\ ms. Subarray readouts are often utilised in order to minimise readout times and obtain the short integrations, which number range between 10 000 to 20 000. The full frame field of view under normal circumstances for AstraLux Norte is $\approx 24'' \times 24''$, and for AstraLux Sur $\approx 15.7'' \times 15.7''$.
The AstraLux observations are mainly performed in the SDSS $z'$-band but in some cases also in the $i'$-band. We primarily use the $z'$-band images for the astrometric analysis here because they are of higher quality and affected by less atmospheric refraction in comparison to the $i'$-band \citep{Bergfors 2010}.

Data reduction for AstraLux is formed with the pipeline that is described by \citet{Hormuth}, identical to previous AstraLux surveys \citep[e.g.][]{Bergfors 2010, Janson 2012, Janson 2014, Janson 2016}. Several Lucky Imaging outputs are produced by the reduction pipeline and we select $10\%$ of the frames which has the best seeing for further analyses, which provides a good trade-off between resolution and sensitivity.

As the system is quite compact (separations of $\lesssim 1''$), we use Point Spread Function (PSF) fitting to determine the relative astrometric positions and angles between the components. Normally, we would use a single star as a reference PSF to iteratively produce binary configurations that minimise the squared residuals of the fit. However, the later three observed epochs taken between 2015 and 2016 with the AstraLux Sur camera are affected by vibrational motions of the telescope, which causes the components to appear stretched out. In order to obtain good enough astrometric measurements we apply the primary component A as a reference PSF. This works, since A is sufficiently separated from the BC pair so that the PSFs are isolated, and we assume the stretching of the image to behave similarly for all three components. This technique provides us with smaller residuals by typically one third compared to convolving a single star PSF with a stretched out Gaussian or earlier PSF fitting by \citet{Janson 2012}.

Calibration of the astrometric measurements is done by comparing AstraLux and Hubble Space Telescope (HST) observations of the Orion Trapezium Cluster and M15 from \citet{McCaughrean, van der Marel} as reference astrometry. The astrometric measurements are shown in Table~\ref{tab:astrometry} together with the angle of true North and the pixel scale for the AstraLux observations. Astrometric errors listed are determined from standard mean deviations of statistical measurements added quadratically with the calibration uncertainties presented in the same table.

\begin{table*}[t]
{
\caption{Collective astrometric measurements of J1036 B and C. The Positional Angle in the third column have been corrected for True North. The last two columns show the deviation from the orbital fit, seen in Section~\ref{sec:BC_orbit}. }
\begin{tabular}{lccccccc}
\hline
 Date & $d$ & P.A.& Pixel Scale & True North & Reference & $\vert\Delta d\vert/\sigma_d $ & $\vert \Delta \rm{PA} \vert / \sigma_{\rm PA} $ \\
  & [mas] & [deg] & [mas/pixel] & [deg] & & & \\
\hline
2006 May 18 &  $189 \pm 2$ &  $-49.38 \pm 0.14$ & - & - & \citet{Daemgen} & <0.1  & 0. 17 \\
2008 Jan 13 &  $163 \pm 4$ & $-13.48 \pm 1.07$ & $23.58 \pm 0.15$ & $ -0.32 \pm 0.18$ & Astralux N & 0.25 & 0.34 \\
2008 Nov 16 &  $150 \pm 3$ & $11.50 \pm 0.63$ & $23.68 \pm 0.01$ & $0.24 \pm 0.05$ & Astralux N & <0.1 & 0.86 \\
2009 Feb 15 &  $145 \pm 3$ & $18.18 \pm 0.68$ & $23.55 \pm 0.17$ & $0.22 \pm 0.20$  & Astralux N & 0.67 & 0.99 \\
2015 Mar 2 &  $187 \pm 4$ & $-43.14 \pm 0.65$ & $15.23 \pm 0.13$ & $2.87 \pm 0.26$  & Astralux S & 0.25 & 1.74 \\
2015 Nov 27 &  $176 \pm 3$ & $-25.32 \pm 0.75$ & $15.20 \pm 0.13$ & $2.40 \pm 0.36$  & Astralux S & 0.67 & 1.09 \\
2016 May 17 &  $164 \pm 3$ & $-14.35 \pm 0.67$ & $15.27 \pm 0.06$ & $3.04 \pm 0.16$ & Astralux S & 0.33 & 0.27 \\
 \\
\hline
\label{tab:astrometry}
\end{tabular}
}
\end{table*}

\subsection{Age, distance and group membership}

Knowing the age of the stellar components is imperative when placing young stars in a mass-luminosity diagram, especially at very young ages when the characteristics of the star changes drastically \citep{Baraffe 1998, Baraffe 2015}. Previous studies of J1036 assumes an age between 35 -- 300\ Myr, based on young moving group association and stellar evolutionary models \citep{Daemgen, Shkolnik 2009, Shkolnik}. 
\citet{Shkolnik} place J1036 in the Ursa Major Moving Group (UMa MG) with a $42\%$ membership probability with the velocity space parameters $UVW_{\rm J1036} = (15.1, 2.4, -7.4) \pm (1.2, 1.0, 0.7)\ $km/s.  \citet{Klutsch} obtain a higher membership probability of $94\%$ by using a more rigorous statistical approach and different $UVW$ for the UMa MG. The discrepancy in membership probability is likely to originate from the different kinematics for the UMa MG. Comparing with the nucleus of the UMa MG estimated by \citet{Mamajek} as $UVW_{\rm UMa nucleus} = (15.0, 2.8, -8.1) \pm (0.4, 0.7, 1.0)\ $km/s suggests that the system is a bona fide member.
The estimated parallax distance to J1036 of $20.1 \pm 2.0$ is also consistent with the distance to the nucleus of the UMa MG of $\approx 25\ $pc.

X-ray luminosities can also be used as an indicator for youth among stellar populations, where young, low-mass stars that have yet to move onto the main sequence usually exhibit more X-ray emission than their older counterparts \citep{Palla}. Using the X-ray luminosity from the ROSAT catalogues \citep{Voges B, Voges F}, \citet{Shkolnik 2009} set an upper limit for the age of J1036 as $\sim 300\ $Myr. This upper age limit is consistent with earlier estimates of the age of the UMa MG of 300\ Myr by \citet{Soderblom}. However, the age of the moving group has since been discussed and revised several times and the now more accepted age is 500\ Myr \citep[e.g.][]{King 2003, King 2005, Brandt}. \citet{Jones} use interferometric measurements and modelling of A type stars in Ursa Major to determine a more precise and consistent age of $414 \pm 23\ $Myr. Nevertheless, the X-ray and UV emissions from J1036 imply signs of youth by showing similar trends as other low-mass stars of comparable spectral type in the UMa MG (e.g. 2MASS J11240434+3808108).  Furthermore, the X-ray emission exhibited by J1036 is also similar to that of systems associated with other young moving groups of different ages ranging between 25 and 600\ Myr, for example 2MASS J03223165+2858291 in the Hyades moving group and GJ 490 B in the TucHor moving group \citep[or Pleides, see][]{Klutsch}.

Distance measurements for J1036 vary from 7\ pc up to 20\ pc, which plays a key role when determining the absolute magnitude of the system. Earlier unresolved spectroscopic measurements by \citet{Riaz} indicate a distance of 7\ pc, whereas \citet{Lepine 2013} obtain both spectroscopic and photometric parallaxes closer to $11.5\ $pc. The resolved photometry by \citet{Daemgen} shows a greater distance of 19.6 pc, and \citet{Shkolnik 2009, Shkolnik} measure the trigonometric parallax to be $20.1\ $pc. As both the photometric and trigonometric parallax display similar distances and are resolved, we deem them to be more reliable and adopt the distance of $20.1 \pm 2.0\ $pc for our analysis of the triplet system J1036. However, the trigonometric parallax is obtained from only four epochs and may therefore not be very well constrained. 


\section{Orbit determination \& mass estimates}\label{sec:orbits}

\subsection{The orbit of J1036 BC} \label{sec:BC_orbit}

In Table~\ref{tab:astrometry} we list the observations and astrometric measurements that we fit orbit models to in order to obtain the orbital parameters for the BC pair. The procedure we follow is described in more detail in \citet{Koehler 2008, Koehler 2012} and can be summarised as a grid-search in periastron $T_0$, period $P$ and eccentricity $e$. We determine Thiele-Innes elements for each grid-point by a linear fit to the astrometric data utilising singular value decomposition. The inclination $i$, angle between node and periastron $\omega$, the position angle of the line of the nodes $\Omega$ and the semi-major axis $a$ are then computed from the Thiele-Innes elements. The minimum we find in the grid-search is then used to improve the estimate for $T_0$ by refining the grid step size to less than one day.

We further improve the grid-search by  simultaneously fitting all seven parameters with a Levenberg-Marquardt $\chi^2$ minimisation algorithm \citep{Press 1992}. The technique depends heavily on the chosen starting values and certain orbits may be biased given poor initial conditions and orbital coverage. Much of the orbit of J1036 B-C remains uncharted and our observations only cover $\lesssim 20\% ~ ( \lesssim 72\ ^{\circ})$ of the entire orbit. Nevertheless, since orbital monitoring began of the system, a full period has been carried out and we can constrain the period of the BC pair to be $\approx 8.4\ $yrs. We compute $\chi^2$ with the formula
$$
\chi^2 = \sum_{i} \left( \biggl( \frac{d_{\rm{obs},i} - d_{\rm{mod},i}}{\sigma_{d,i}} \biggr)^2 + \biggl( \frac{\rm{PA}_{\rm{obs},i} - \rm{PA}_{\rm{mod},i}}{\sigma_{\rm{PA},i}} \biggr)^2 \right),
$$
where PA and $d$ are the position angles and separation, $\sigma_{\rm PA}$ and  $\sigma_d$ are their respective errors. We mark the observations and model predictions by the suffixes 'obs' and 'mod' respectively.

\begin{figure}[hbtp]
  \centering
  \includegraphics[width=\columnwidth]{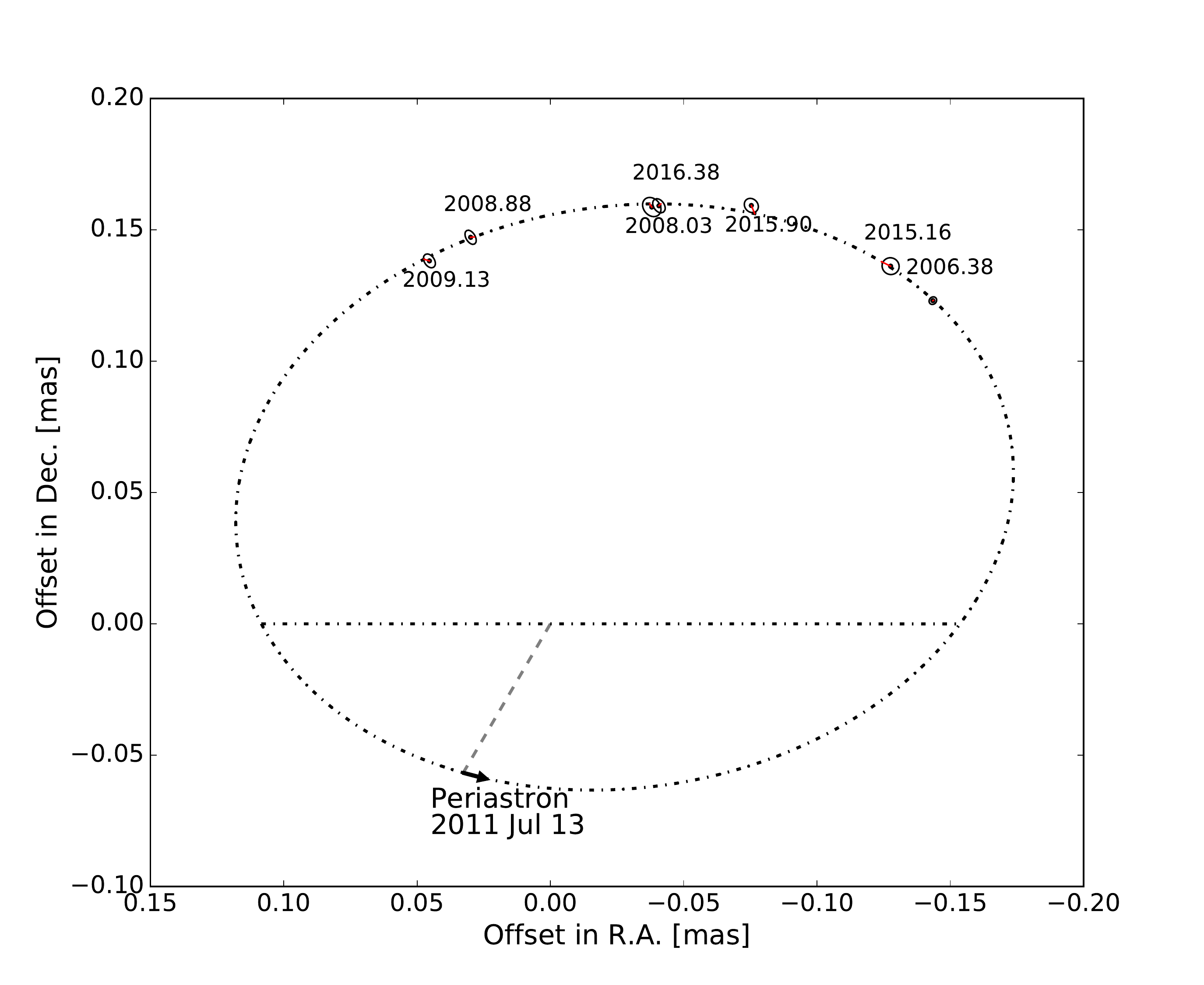}
  \caption[]{\label{fig:BC} %
    Best-fit orbit of C around B in the rest frame of B. The ascending and descending nodes are marked by the dash-dotted line and the periastron by the dashed line. The ellipses mark the errors of the observed positions, with red lines connecting the observed positions with the calculated orbit. The orbital motions is counter-clockwise, as depicted by the arrow.
  }
\end{figure}

The resulting orbital parameters from the best-fit orbit are listed in Table~\ref{tab:BC} and the calculated orbit shown in Figure~\ref{fig:BC}. The differences between the observed astrometric data points and the orbit fit are listed in the last two columns of Table~\ref{tab:astrometry}. We use Kepler's Third Law \citep{Kepler} to compute the system mass $M_{\rm B + C}$ as
$$
M = \frac{a^3}{ P^2},
$$
where $a$ is the major-semi axis in mas and $P$ the period in years.  We translate the system mass into units of solar masses by converting the angular separation to a linear separation by multiplying it by the distance 20.1\ pc and obtain $M_{\rm B+C} = 0.482 \pm 0.145\ M_\odot$. 

We obtain a reduced $\chi^2 = 1.0$, which is expected of a good fit.
Errors of the orbital elements are obtained by examining the minimum of the $\chi^2$ function,  and the points where $\chi^2 = \chi^2_{\rm min} +1$ correspond to the uncertainty for each parameter. By perturbing one parameter at a time (e.g. $T_0$) away from the minimum while optimising all other parameters, we find good estimates for the value of $T_0$ where $\chi^2 = \chi^2_{\rm min}+1$. This procedure is repeated to find errors for the other parameters.  For more details, see \citet{Koehler 2012}.
This error is small in comparison to the error from the distance measurements which dominates the final error of the system mass.

\begin{table}
\caption{Best orbital solution parameters for B--C}
\label{tab:BC}
\renewcommand{\arraystretch}{1.3}
\begin{center}
\begin{tabular}{lr@{}l}
\noalign{\vskip1pt\hrule\vskip1pt}
Orbital Element				& \multicolumn{2}{c}{Value} \\
\noalign{\vskip1pt\hrule\vskip1pt}
Periastron date $T_0$			& $2455756$ & $\,^{+   3}_{  -26}$\\
						& (2011 Jul 13)\span\\
Period $P$ (years)                              & $   8.41$ & $\,^{+0.04}_{-0.02}$\\
Semi-major axis $a$ (mas)                       & $  161$ & $\,^{+2}_{-2}$\\
Semi-major axis $a$ (AU)                        & $    3.2$ & $\,^{+0.3}_{-0.3}$\\
Eccentricity $e$                                & $  0.460$ & $\,^{+0.005}_{-0.007}$\\
Argument of periastron $\omega$ ($^\circ$)      & $     68$ & $\,^{+   3}_{  -2}$\\
P.A. of ascending node $\Omega$ ($^\circ$)      & $   90.0$ & $\,^{+0.4}_{-0.3}$\\
Inclination $i$ ($^\circ$)                      & $   45.4$ & $\,^{+3.8}_{-0.9}$\\
System mass $M_{\rm{B}+\rm{C}}$ ($\rm mas^3/year^2$)          & $59330$ & $\,^{+2000}_{-2000}$\\
Distance error ($M_\odot$)	& $	$ & $\pm   0.144$ \\	
System mass $M_{\rm{B}+\rm{C}}$ ($M_\odot$)                   & $    0.482$ & $\,^{+0.145}_{-0.145}$\\

reduced $\chi^2$				& $    1.0$\\

\noalign{\vskip1pt\hrule\vskip1pt}
\end{tabular}
\end{center}
\end{table}



\subsection{The orbit of J1036BC around A}

With the combined BC system mass known, we only require the mass ratio $q$ in order to determine their individual masses. We calculate the mass ratio by using that the centre of mass (CM) between B and C is on an orbit around A, as well as that the BC pair are orbiting their own mutual CM. Although the CM between B and C is not directly observable, it is located on a line between them so that the fractional mass $f = q/(1+q)$ is a constant fraction of the separation of B and C from its distance to B \citep{Heintz}.

\begin{figure}[hbtp]
  \centering
	\includegraphics[width=\columnwidth]{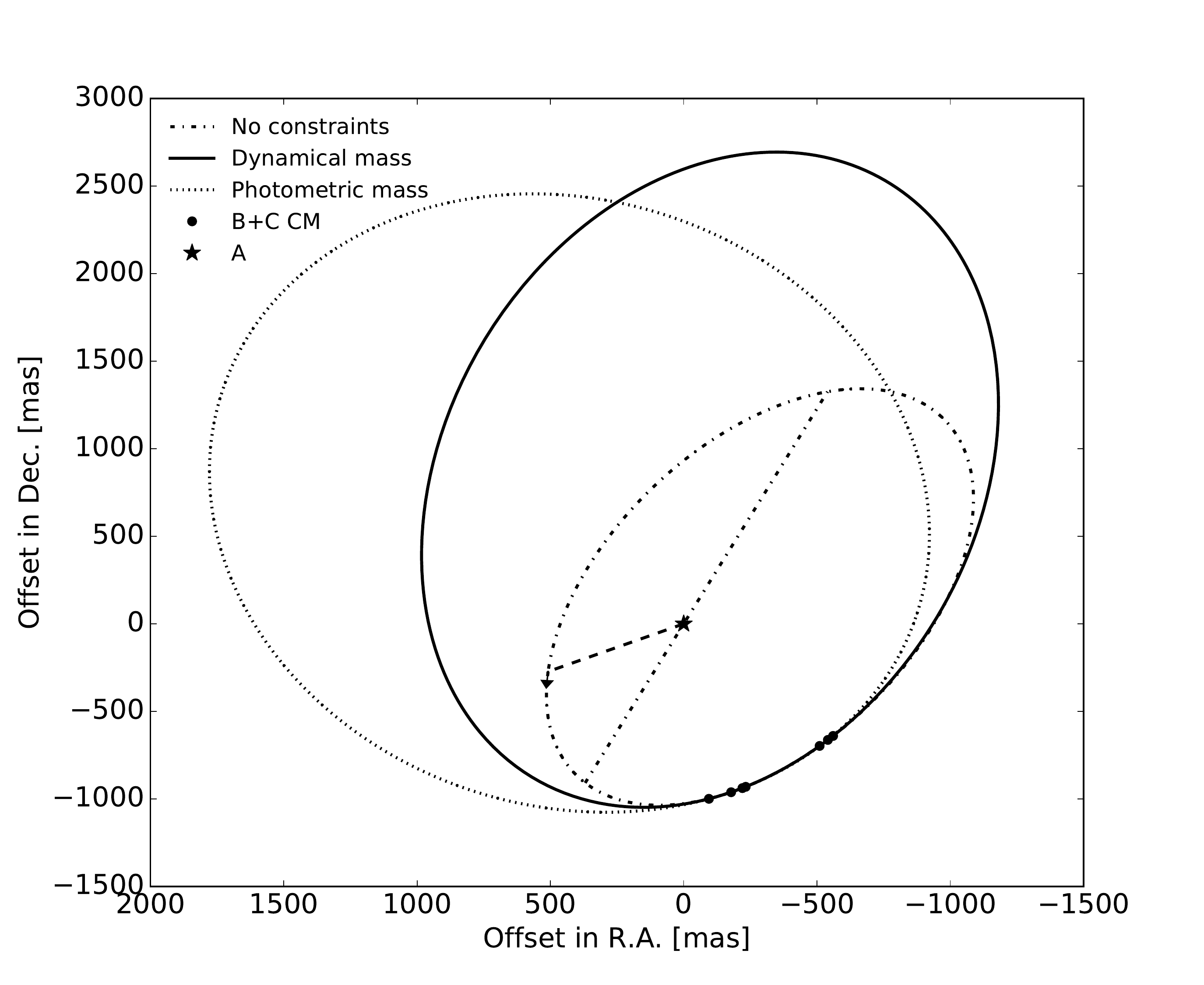} \
	\includegraphics[width=\columnwidth]{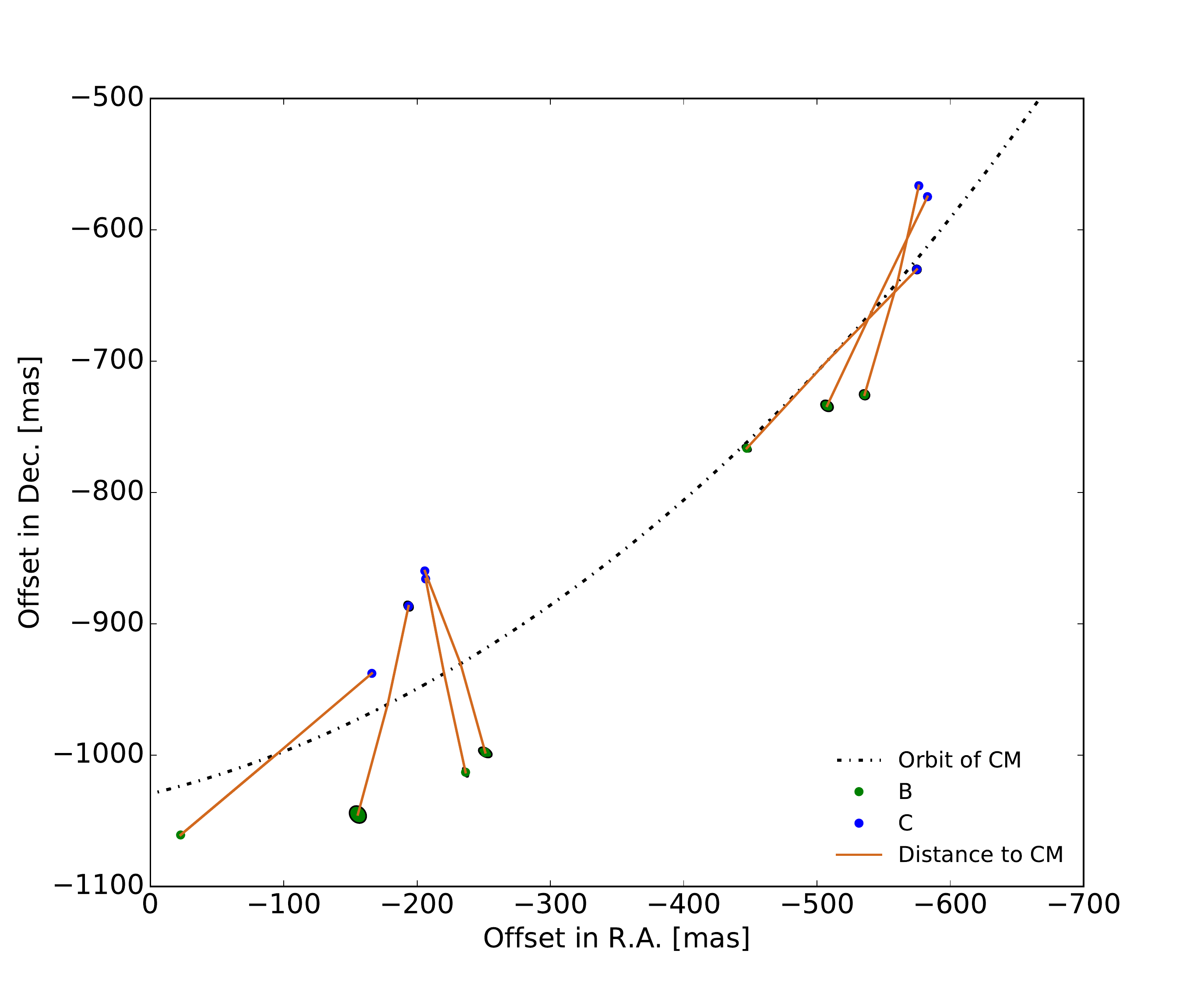}
  \caption[]{\label{fig:outer} %
		Motion of J1036 B and C centre of mass in the reference frame of J1036 A. Upper plot shows the full estimated period of the centre of mass between B and C around A for our fit using no mass constraints (dashed), our dynamical mass of the system as $M_{\rm est, dyn} = 1.0 \pm 0.3\ \ M_\odot$ (solid) and the photometric mass estimate of the system as $M_{\rm est, phot} = 0.65 \pm 0.2 \ \ M_\odot$. The lower plot depicts the same as the above but zoomed in on the observed data points where the ellipsis depicts the errors of the B (green) and C (blue) component positions. The brown lines connect B and C to the computed centre of mass. Only $\approx 35\ ^{\circ}$ of the full orbit has been covered and thus may not be well represented here.
  }
\end{figure}

The method we use to estimate the mass ratio of J1036 B and C is the same as described in \citet{Koehler 2008, Koehler 2012, Koehler 2016} for the triple systems T Tauri and LHS 1070.  We carry out another grid-search, this time for the A-BC binary with the additional dimension of $f$, which only refers to the BC pair. The resulting outer orbit is shown in the lower plot of Figure~\ref{fig:outer}. From the relative positions of the components we derive the mass fraction $q = 1.00 \pm 0.03$, meaning that components B and C are of equal mass. This result corresponds well with the equal magnitudes measured in all observed bands. Since the total mass of the BC pair is estimated to be $M_{\rm B+C} = 0.48\ M_\odot$, we conclude that the individual masses are $M_{\rm B} = 0.24\ M_\odot$ and $M_{\rm C} = 0.24\ M_\odot$. Earlier studies of the binarity of M-dwarfs by \citet{Daemgen} suggest that near-equal mass companions are common\footnote{In their paper, \citet{Daemgen} count J1036 twice; first as the A-BC pair, then also the B--C pair.}, but this may be one of the few cases where the precision of the ratio is so close to unity.

The orbit of the  BC pair around A is only partially covered and a naive best-fit orbit yields a period of $\approx$100 years, shown by the dashed line in the upper plot of Figure~\ref{fig:outer}. The corresponding system mass exceeds $2\ M_\odot$, which seems too high and very unlikely in comparison to the dynamical mass estimate of the other components B and C and their relative measured brightness. Nevertheless, following the same approach as \citet{Koehler 2013} we can constrain the estimated mass of the orbital fit, and make our prediction of the orbit with an already predetermined mass. This procedure does not provide us with any new information of the mass, as what we put in is the value that will come out of the fit. The orbital fitting procedure is the same as described above with an additional term to the computation of the $\chi^2$ of the form 

$$\left( \frac{M_{\rm mod} - M_{\rm est}}{0.3\ M_\odot} \right)^2 $$

where the orbit model system mass is $M_{\rm mod}$ and the estimate of the mass we plug in is $M_{\rm est}$. We run the orbital fit for 3 scenarios: without any mass constraint, with the dynamical mass estimate as a constraint, and with a photometric mass estimate as a constraint. We estimate the dynamical model mass as $M_{\rm est, dyn} = 1.0 \pm 0.3\ M_\odot$ by using our earlier estimate of the mass of B+C as $0.48\ M_\odot$. We reason that A should be of same order in terms of mass as the combined B+C, even slightly more so as it is brighter than the combined brightness of B$+$C, yet we do not expect the mass of A alone to exceed $0.6\ M_\odot$ based on their relative brightness. In order to give a mass estimate purely based on photometry we adopt the mass-luminosity relation of \citet{Benedict} as

\begin{multline*} 
 M = C_0 + C_1 (K_{\rm mag} - x_0) + C_2 (K_{\rm mag} - x_0)^2 + \\
 C_3 (K_{\rm mag} - x_0)^3 + C_4 (K_{\rm mag} - x_0)^4,
\end{multline*}

where $C_0 = 0.2311$, $C_1 = -0.1352$, $C_2 = 0.04$, $C_3 = 0.0038$, $C_4 = -0.0032$, $x_0 = 7.5$ and $K_{\rm mag}$ is the absolute magnitude in the K-passband. By using the observed magnitudes presented in Table~\ref{tab:photometry} for the K-band and the distance of 20.1 pc we obtain a total mass of the system of $M_{\rm est, phot} =0.65 \pm 0.2 M_\odot$ ($ 0.314 + 0.166 + 0.166\ M_\odot$ for components A, B and C respectively).
The uncertainties in the mass constraints for $M_{\rm est}$ are dominated by the error from the distance.

We run our orbital fitting procedure for the three scenarios, shown in the upper plot of Figure~\ref{fig:outer}. The orbital period of B and C around A varies from $\approx$100 years in the 'no constraint' orbital fit,  $\approx 230 \,^{+   10}_{ -30}$ years with our dynamical mass estimate as a constraint, to $\approx 280 \,^{+  20}_{ -20}$ years using the photometric mass estimate. As our primary result, we adopt the orbital fit when the system mass is constrained to $1.0 \pm 0.3 M_\odot$ as our dynamical model suggests and we show the resulting parameters in Table~\ref{tab:outer}.  The resulting parameters obtained from the outer orbit are mainly used for illustrative purposes and to test the mass ratio $q$ between the B and C components.
For all three of the outer orbital fits we obtain a mass ratio of B to C as $q = 1.00 \pm 0.03$, suggesting that the mass ratio is adequately constrained by the astrometric data. We also obtain for all three cases a reduced $\chi^2 = 1.6$, indicating that our uncertainties for the astrometry are underestimated. 
Further orbital coverage with future observations are required to make better estimates of the full outer orbit.

\begin{table}
\caption{Best orbital solution parameters for A--BC when the system mass is constrained by the the B$+$C pair dynamical mass of $0.48 \pm 0.14 M_\odot$.}
\label{tab:outer}
\renewcommand{\arraystretch}{1.3}
\begin{center}
\begin{tabular}{lr@{}l}
\noalign{\vskip1pt\hrule\vskip1pt}
Orbital Element				& \multicolumn{2}{c}{Value} \\
\noalign{\vskip1pt\hrule\vskip1pt}
Periastron date $T_0$                        & $2452302$ & $\,^{+ 223}_{-1156}$\\	
                                                & (2002 Jan 28)\span\\
Period $P$ (years)                              & $    234$ & $\,^{+   10}_{ -29}$\\	
Semi-major axis $a$ (mas)                       & $   1898$ & $\,^{+  53}_{-105}$\\	
Semi-major axis $a$ (AU)                        & $   38$ & $\,{\pm 4}$\\	
Eccentricity $e$                                & $     0.44$ & $\,^{+ 0.03}_{-0.02}$\\	
Argument of periastron $\omega$ ($^\circ$)      & $      9$ & $\,^{+   2}_{  -2}$\\	
P.A. of ascending node $\Omega$ ($^\circ$)      & $    167$ & $\,^{+   6}_{  -3}$\\	
Inclination $i$ ($^\circ$)                      & $     52.5$ & $\,^{+ 0.3}_{-0.4}$\\	
System mass $M_{A+B+C}$ ($\rm mas^3/yr^2$)    & $ 124658$ & $\,^{+ 696}_{-689}$\\	
System mass $M_{A+B+C}$ ($M_\odot$)           & $ 1.01$ & $\,{\pm 0.3}$\\	
Mass ratio $M_{C}/M_{B}$                      & $     1.00$ & $\,{\pm 0.03}$\\	
reduced $\chi^2$				&  & $    1.6$\\

Mass of A $M_A$	 ($M_\odot$) & $	0.53$ & $\,{\pm 0.20}$\\		
Mass of B $M_B$ ($M_\odot$) & $	0.24$ & $\,{\pm 0.07}$\\		
Mass of C $M_C$ ($M_\odot$)  & $	0.24$ & $\,{\pm 0.07}$\\

\noalign{\vskip1pt\hrule\vskip1pt}
\end{tabular}
\end{center}
\end{table}

\section{Evolutionary models} \label{sec:models}

\begin{figure*}[hbtp]
  \centering
  \includegraphics[width=\columnwidth]{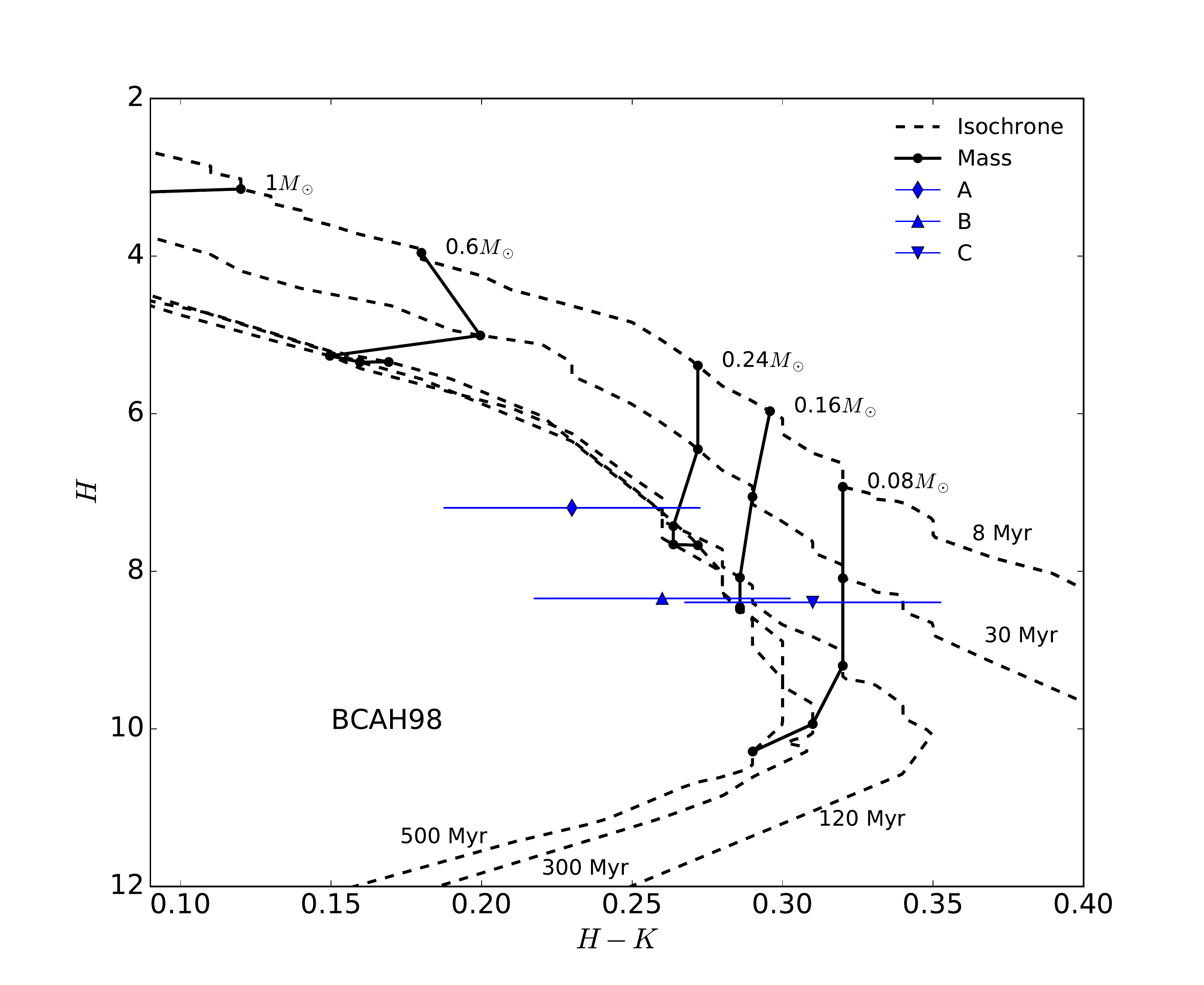}%
  \includegraphics[width=\columnwidth]{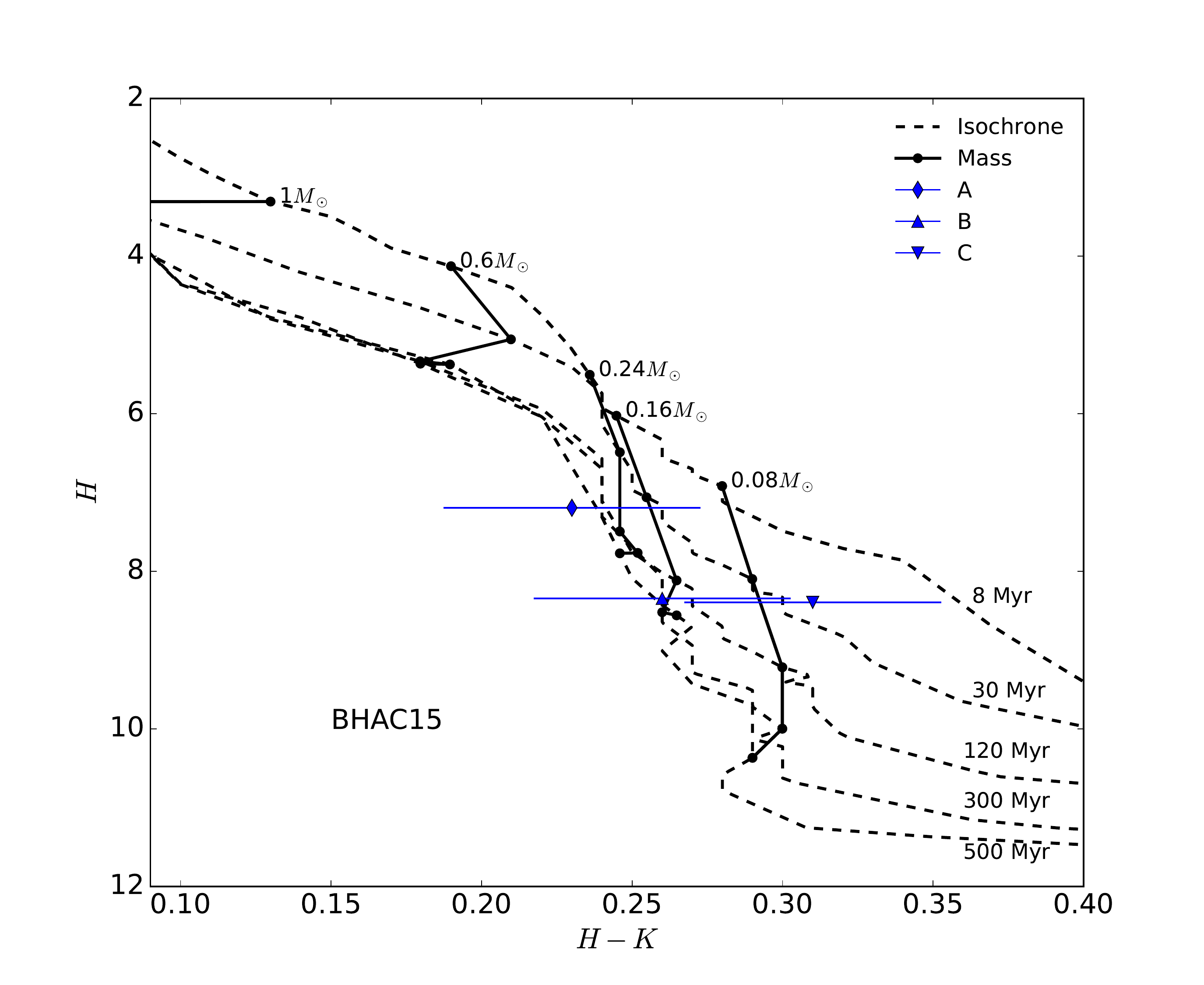}\  
  \includegraphics[width=\columnwidth]{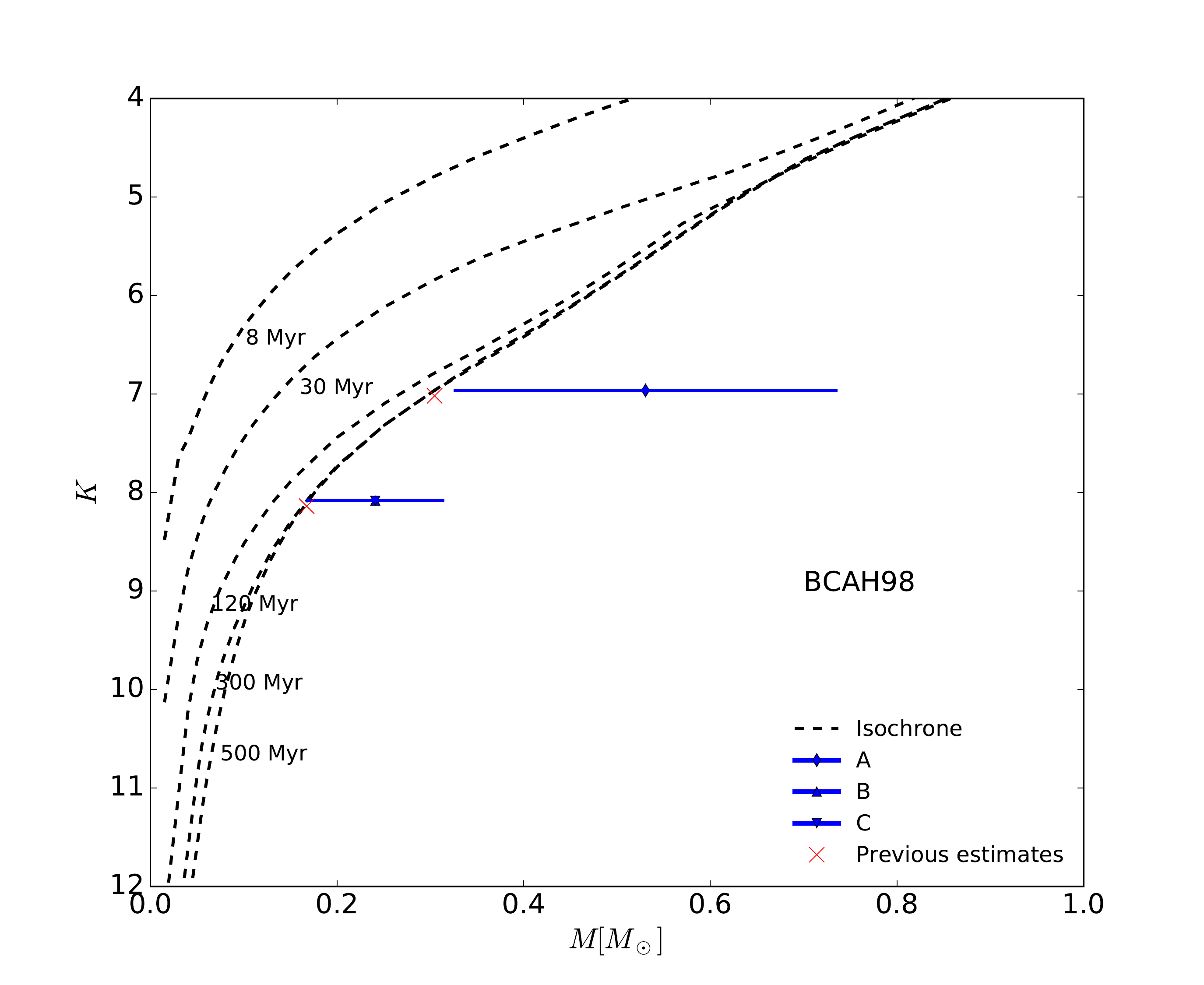}%
  \includegraphics[width=\columnwidth]{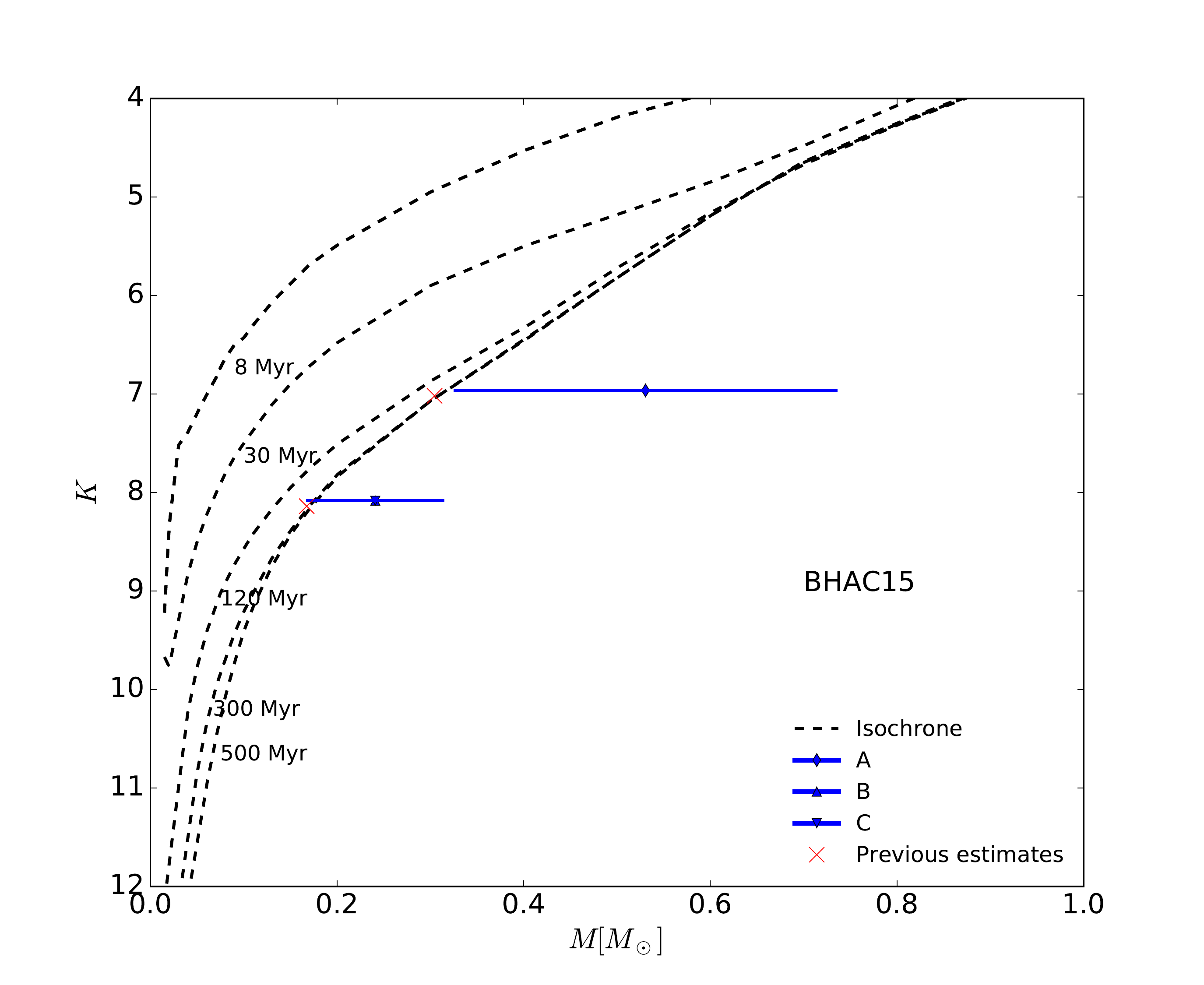}
  \caption[]{\label{fig:isochrones} %
 \citet[{\bf left}]{Baraffe 1998} and \citet[{\bf right}]{Baraffe 2015} stellar evolutionary models with isochrones ranging from $8 -30 -120 - 300 - 500\ $Myr. Earlier models combined with the photometry from \citet{Daemgen} suggest that the BC components of J1036 are $\approx 300$ Myrs old and close to $0.16\ M_\odot$ in mass. More recent models however, do not agree and with the current measurements and would argue for the components being far much younger and less massive, and put in the brown dwarf regime where hydrogen has yet to begun nuclear fusion. Neither model correspond well with our estimates of the individual dynamical mass of B and C of $0.24\ M_\odot$, indicated by the blue markers in the lower plot, which are within $1\ \sigma$ of the models predictions where the uncertainty is dominated by the error from the distance of $20.1 \pm 2.0\ $pc. The red crosses in the lower plot are earlier mass estimates, where the observed brightness has been interpolated with the 300 Myr isochrone. The length of the arms of the red crosses do not represent the errors. 
  }
\end{figure*}

\begin{figure}[hbtp]
  \centering
  \includegraphics[width=\columnwidth]{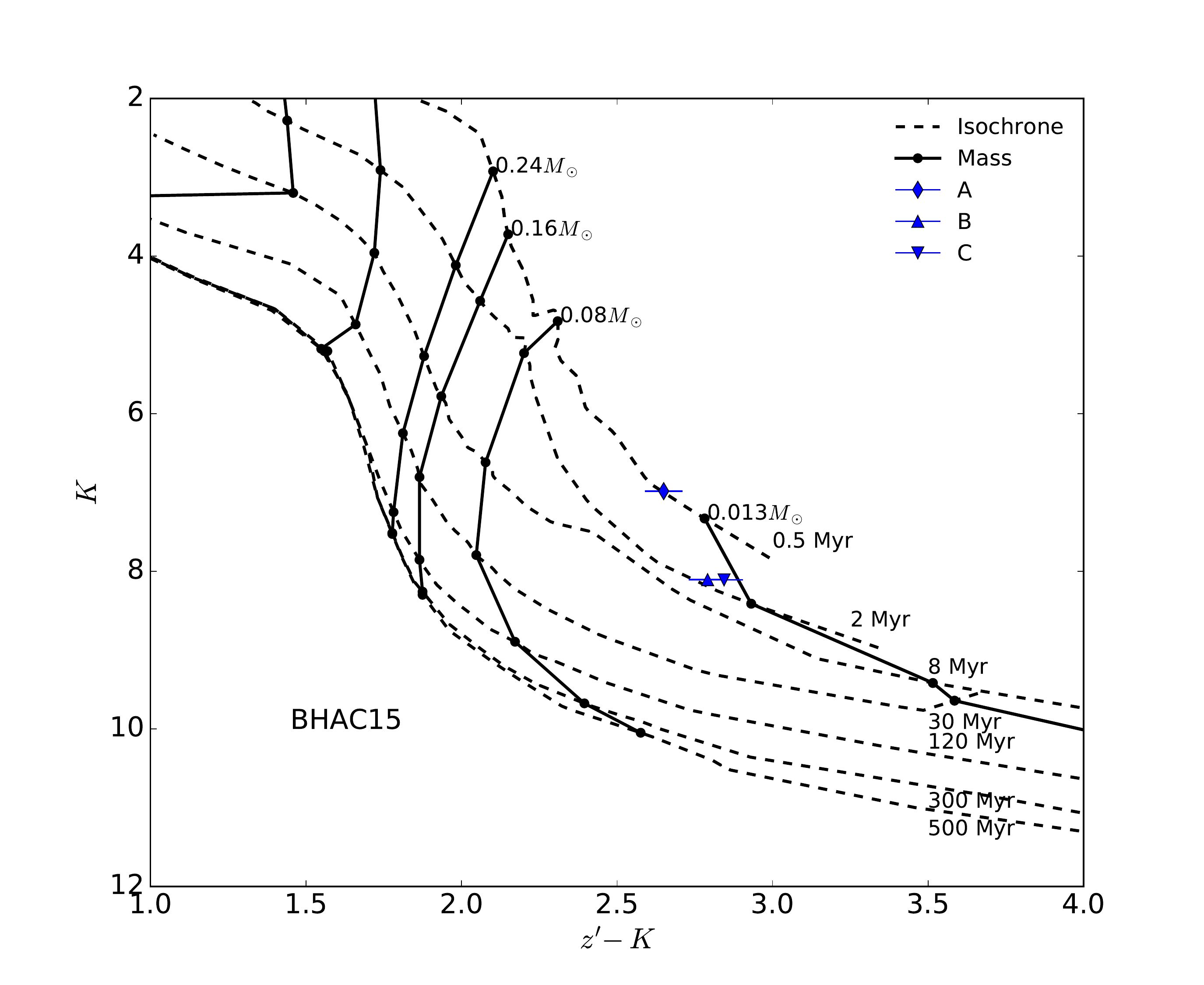}
  \caption[]{\label{fig:ukidss} %
    \citet{Baraffe 2015} stellar evolutionary models with isochrones ranging from 0.5 Myr to 500 Myr plotted with $z'$ and $K$ absolute magnitudes. The observed data of J1036 is depicted by the blue diamond and triangles for the A, B and C components, where we have used the 20.1 pc distance from \citet{Shkolnik}.  Similar to Figure~\ref{fig:isochrones}, the models do not agree with the individual dynamical mass of $0.24\ M_\odot$ for the B and C components. The models also suggest for a very young age unless the distance to the system has been overestimated.
  }
\end{figure}

The dynamical mass estimates here can be used to test how well theoretical stellar evolutionary models can predict the mass, luminosity and age of the M-type stars. From the literature we compile the observed apparent magnitudes for J1036 in the cases where all three components have been resolved, shown in Table~\ref{tab:photometry}. We then compare the observed brightness of J1036 to the evolutionary models of \citet[][referred to as BCAH98 and BHAC15 respectively]{Baraffe 1998, Baraffe 2015}\footnote{\url{http://perso.ens-lyon.fr/isabelle.baraffe/BHAC15dir/}}, shown in Figures~\ref{fig:isochrones} \& \ref{fig:ukidss}. The AstraLux observations only measures relative fluxes between the resolved components, which we combine with unresolved measurements from \emph{SDSS} \citep{Adelman} in order to calculate individually resolved magnitudes for each component. The error bars in Figure~\ref{fig:isochrones} are directly translated from Table~\ref{tab:photometry} and in the case for $H - K$ magnitudes we add the uncertainties for each band quadratically.

\begin{table}[t]
{\scriptsize
\caption{Photometric measurements of J1036.}
\begin{tabular}{ccccc}
\hline \hline
 Bandpass & Component & Magnitude & Uncertainty & Source \\
\hline
H & A & 8.72 & 0.03 & \citet{Daemgen}\\
H & B & 9.86 & 0.03 & \citet{Daemgen}\\
H & C & 9.91 & 0.03 & \citet{Daemgen}\\
K & A & 8.48 & 0.03 & \citet{Daemgen}\\
K & B & 9.60 & 0.03 & \citet{Daemgen}\\
K & C & 9.60 & 0.03 & \citet{Daemgen}\\
\hline
$i'$ & A & 11.70 & 0.15 & \citet{Janson 2012}\\
$i'$ & B & 13.03 & 0.15 & \citet{Janson 2012}\\
$i'$ & C & 13.03 & 0.15 & \citet{Janson 2012}\\
$z'$ & A & 11.13 & 0.05 & \citet{Janson 2012}\\
$z'$ & B & 12.39 & 0.05 & \citet{Janson 2012}\\
$z'$ & C & 12.45 & 0.05 & \citet{Janson 2012}\\
 \\
\hline
\label{tab:photometry}
\end{tabular}\\
}
\end{table}

\citet{Daemgen} interpolates the 300 Myr isochrone BCAH98 model together with the observed absolute magnitudes in order to obtain the individual masses of the components as $M_{\rm A} = 0.29\ M_\odot$, $M_{\rm B} = 0.16\ M_\odot$ and $M_{\rm C} = 0.16\ M_\odot$. We follow the same procedure and show a comparison of the predicted individual mass from several stellar evolutionary models and isochrones in Table~\ref{tab:masses}. The more recent models, seen in the upper right plot of Figure~\ref{fig:isochrones}, do not correspond well with an age of 300 Myr, and the brightness of each component is more in line with the 30 Myr isochrone, which would suggest individual masses for components B and C to be below the Hydrogen burning limit. The lower plots in Figure~\ref{fig:isochrones} illustrate the discrepancy between the dynamical and  the theoretical masses, which also makes it difficult to place an age estimate on J1036 based solely on photometry.

\begin{figure}[hbtp]
  \centering
  \includegraphics[width=\columnwidth]{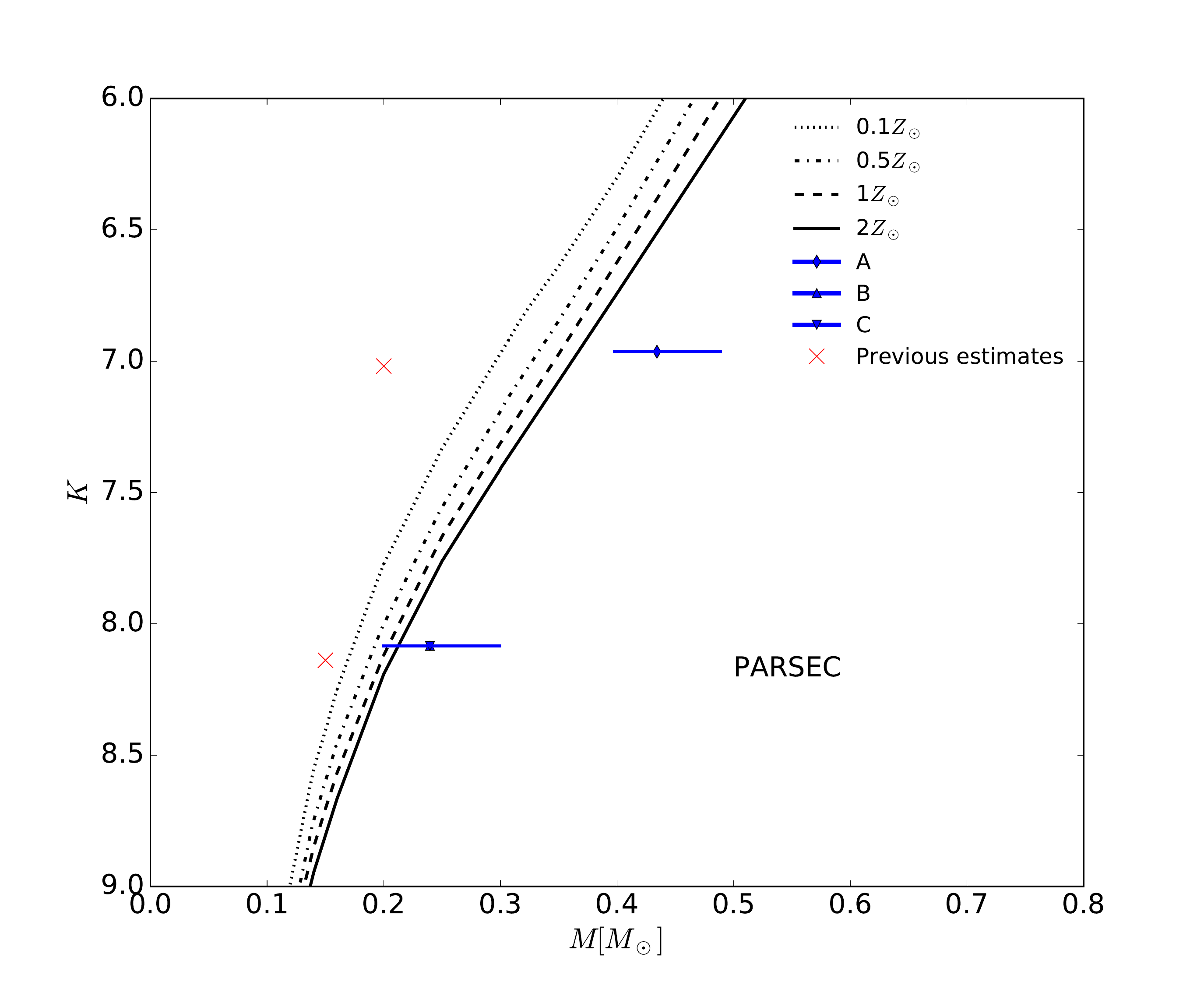}
  \caption[]{\label{fig:CMD} %
    Evolutionary tracks of \citet{Bressan} with four tracks for different metallicities for the 500 Myr isochrone. Our dynamical mass estimates are depicted by the blue triangular and diamond markers and the earlier photometric estimates from \citet{Daemgen} by the red crosses (the arm length does not represent an error estimate for the cross).  Solar metallicity is shown by the dashed line and the average metallicity of the Ursa-Major moving group is somewhere between the dashed-dotted and the solid line.
  }
\end{figure}

We also compare our dynamical mass estimates and the observed brightness with the \texttt{PARSEC}\footnote{\url{http://stev.oapd.inaf.it/cgi-bin/cmd}} isochrone evolutionary models of different metallicities from \citet{Bressan}, shown in Figure~\ref{fig:CMD}.  We note that the metallicity effects are too subtle to explain the discrepancy, but also that the both the dynamical and BHAC15 500 Myr isochrone mass estimates are dissimilar from the average metallicity of the UMa MG, which is supposedly near solar \citep{Boesgaard}.


We further analyse the discrepancy with regards to the $z'$-band, and show the even more pronounced mass discrepancy in Figure~\ref{fig:ukidss} where we plot the $z' - K$ magnitude difference, and include two more isochrones of 2\ Myr and 0.5\ Myr to further illuminate the discrepancy. The evolutionary tracks place the observed brightness of the components of J1036 with the adopted distance of 20.1\ pc between the ages of 0.5-2\ Myr,  which would yield masses below $20\ M_{\rm jup}$.

Since the biggest uncertainty in the estimates of the dynamical mass and the photometry stems from the distance to the system, we probe the consequences of adopting another distance measurement than the trigonometric parallax from \citet{Shkolnik}. If we adopt the distance of 7\ pc to our calculations we obtain dynamical masses of B and C to be below the Hydrogen burning limit. This is a particular interesting result, as this would again place the components B and C on the 300\ Myr isochrone in the more recent evolutionary models, which also happens to agree with the very low mass estimate. However, the distance of 7\ pc is based on unresolved spectroscopic measurements and thus considered to be less reliable than the trigonometric parallax, the latter which is also in broad agreement with the resolved photometric distance of 19.6 pc from \citet{Daemgen}.

We further probe at what distance to the system the theoretical and dynamical mass estimates would coincide. This distance corresponds to $12.3 \pm 0.2$ pc for the 500 Myr BHAC15 isochrone, which is very similar to the unresolved photometric parallax distances measured by \citet{Lepine 2013} and \citet{ Shkolnik} of $11.8 \pm 3.5$pc and $12 \pm 3$ pc respectively. However,  that would imply individual masses of $M_{\rm A} = 0.12 \pm 0.06 M_\odot$ and $M_{\rm B (C)} = 0.06 \pm 0.03 M_\odot$, thus placing the BC pair in the brown dwarf mass regime below the hydrogen burning limit, which is a highly unlikely scenario when considering their measured spectral types of M4.5$\pm $0.5.


\begin{table}[t]
{
\tiny
\centering
\caption{Mass predictions from theoretical models and $K$-band photometry. Errors for models include measured photometry and distance uncertainties.}
\begin{tabular}{lccc}
\hline \hline
 Model & Age [Myr] & A [$M_\odot$] & B or C [$M_\odot$]  \\
\hline
BHAC15 & 8  &$ 0.063 \pm 0.011$ & $0.023 \pm 0.003$ \\
BHAC15 & 30 & $0.144 \pm 0.021$ & $0.067 \pm 0.010$  \\
BHAC15 & 120 & $0.285 \pm 0.036$ & $0.138 \pm 0.020$ \\
BHAC15 & 300 &$ 0.317 \pm 0.034$ & $0.175 \pm 0.019$  \\
BHAC15 & 500 & $0.317 \pm 0.033$ & $0.178 \pm 0.019$  \\
\hline
BHAC98 & 8 & $0.061 \pm 0.010$ & $0.022 \pm 0.004$  \\
BHAC98 & 30 & $0.140 \pm 0.021$ & $0.064 \pm 0.010$ \\
BHAC98 & 120 & $0.273 \pm 0.037$ & $0.132 \pm 0.019$  \\
BHAC98 & 300 & $0.304 \pm 0.035$ & $0.167 \pm 0.018$  \\
BHAC98 & 500 & $0.304 \pm 0.034$ & $0.169 \pm 0.018$  \\

\hline
PARSEC ($Z_\odot$) & 500 & $0.367 \pm 0.033$ & $0.212 \pm 0.023$ \\ 

\hline
\citet{Benedict} & - & $0.314 \pm 0.037 $& $0.166 \pm 0.019$ \\

\hline
\hline
Dynamical & - & $0.531 \pm 0.204$ & $0.241 \pm 0.072$  \\

 \\
\hline
\label{tab:masses}
\end{tabular}\\
}
\end{table}

\subsection{Spectral analysis}\label{sec:SED}

\begin{figure*}[hbtp]
  \centering
  \includegraphics[width=\columnwidth]{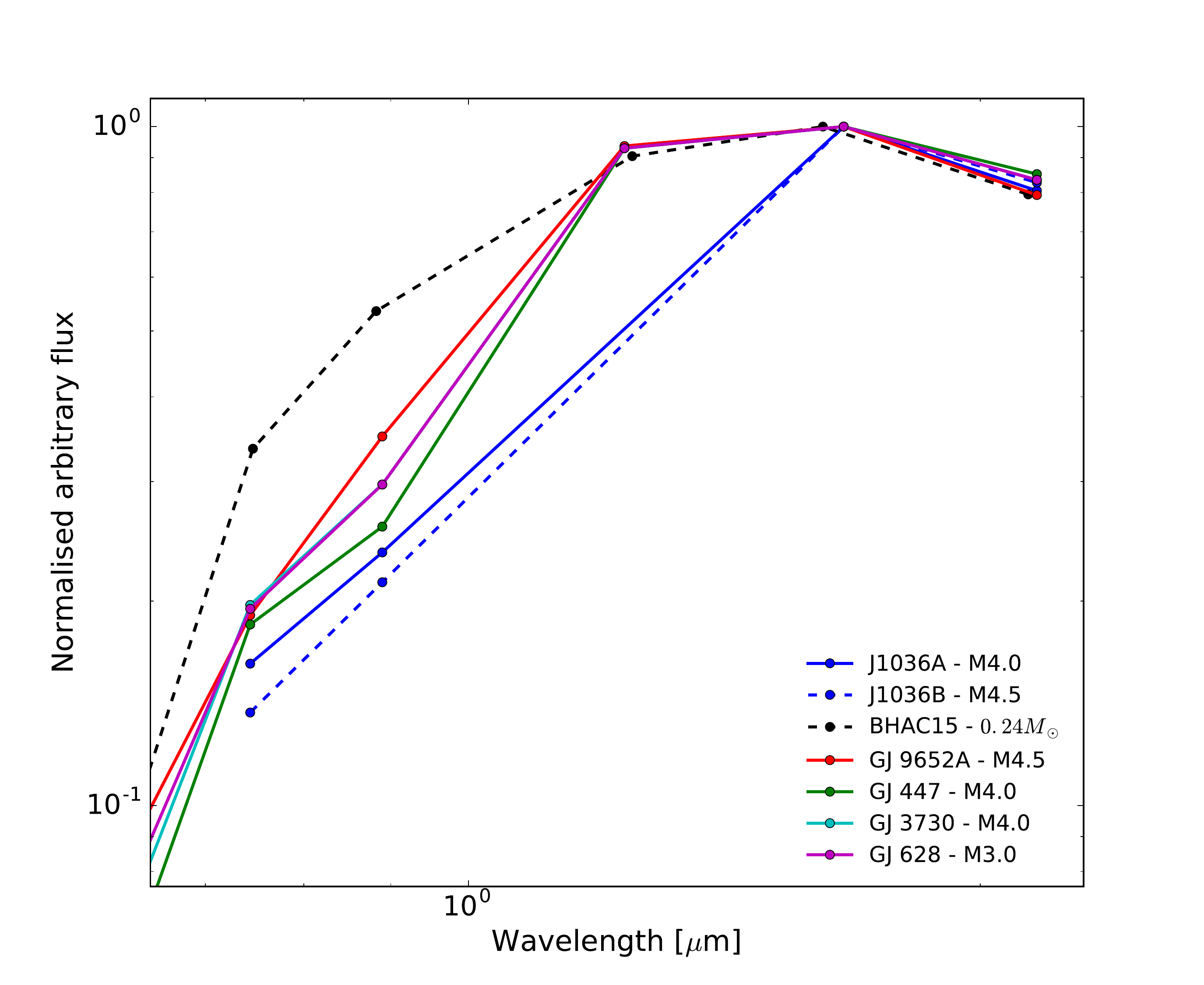} 
  \includegraphics[width=\columnwidth]{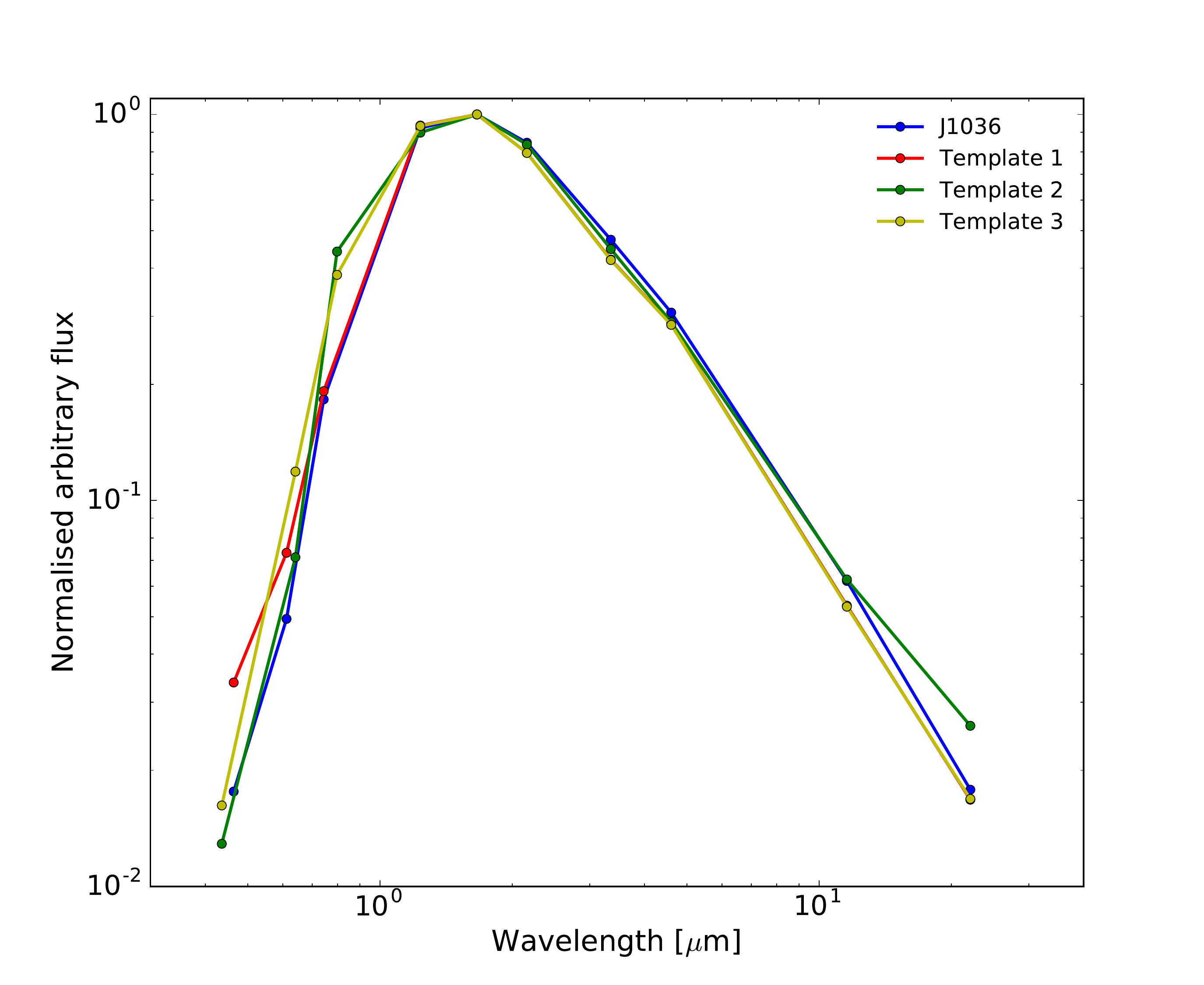}
  \caption[]{\label{fig:SED} %
    Spectral Energy Distribution of J1036 plotted together with several reference stars of the different spectral types \citep{Shkolnik}. The left plot displays the A and B components of J1036 as the blue solid and dashed line respectively. We also include a dashed black line representing the \citet{Baraffe 2015} 500 Myr isochrone model for the same mass as estimated for J1036B of $0.24\ M_\odot$. The major differences in shape between the lines stems from the incompleteness of observational data of J1036 in the $1.0-1.6\ \mu$m wavelength region. There is also a hint of reddening in J1036 as the magnitude seems to drops off at lower wavelengths. 
    The right plot displays the unresolved J1036 system plotted together with three composite template systems of reference stars of the same spectral types as J1036ABC. The bands used in the SED are $g'r'i'JHKw_1w_2w_3w_4$ for J1036 and Template 1 and bands $BRIJHKw_1w_2w_3w_4$ for Template 2 and Template 3.  See the text in Section~\ref{sec:SED} for further information on the three composite templates. The templates fit well with the expected spectral types of J1036, and there are no signs of extreme reddening from the unresolved J1036 system. All lines have been normalised to their respective maximum value, which is represented here by the $H$ band magnitude for all cases.
  }
\end{figure*}

A Spectral Energy Distribution (SED) fit can be used to classify a source by assessing over which wavelength range most of the energy is distributed. Here, we make an SED fit to J1036 and compare it to other stars of various spectral types to test whether there could be some hint towards a reddening in the system which could indicate the need for extra extinction corrections. We compare the resolved observations from Table~\ref{tab:photometry} with other stars of various spectral types from \citet{Shkolnik} that have resolved photometry in similar bands, shown in the left plot of Figure~\ref{fig:SED}. We also include a comparison to the \citet{Baraffe 2015} 500 Myr isochrone track for a star with the same mass as our dynamical mass estimate for J1036B of $0.24\ M_\odot$.


Additionally, we make an SED plot of the unresolved J1036 triple system where we use photometric data from \emph{APASS, 2MASS} and \emph{WISE}, shown in the right plot of Figure~\ref{fig:SED}. We compare the unresolved SED of J1036 with template SEDs that we construct by combining the absolute magnitudes of one reference star of spectral type M4.0 with two of spectral type M4.5 in order to resemble the J1036 system configuration. For Template 1 we combine the stars GJ3730 with two GJ9652A, for Template 2 we combine NLTT22741 with two 2MASS J23500639+2659519 and for Template 3 we combine G 68-46 with two GJ9652A. Template 1 uses the same filters as we obtained for J1036; \emph{SLOAN}$g' r' i'$,  \emph{2MASS} $JHK$ and \emph{WISE} $w_1,w_2,w_3,w_4$. For Template 2 and Template 3 we employ $BRI$ photometric data points instead of the $g'r'i'$, as those filters were not always available for our reference stars. In the cases where a band is missing for our reference stars, typically the $z'$ band in this case, we use the synthetic photometric data from \citet{Pickles} instead.

Both plots in Figure~\ref{fig:SED} suggest a modest reddening of J1036. Given the observed inclination of the orbital plane of B and C around each other and around A, it is unlikely that this reddening is due to all components in the system to possess edge-on discs that would make them appear more red. The small separation between B and C of $3.2 \pm 0.3$ also makes it difficult to harbour substantial circumstellar discs.

We recognise that in our SED analysis not all of the reference stars used have their spectral types very well constrained, and that some of their distances remain ambiguous.


\section{Summary \& conclusion} \label{sec:summary}

From new relative astrometric measurements of  2MASS J10364483+1521394 collected in 2015-2016, we are able to derive a new model for the orbit of the B and C components around each other. The orbital period is well constrained to $P = 8.41^{+0.04}_{-0.02}\ $years. We estimate the dynamical mass of the BC pair to be $M_{\rm{B}+\rm{C}} = 0.482 \pm 0.145\ M_\odot$, and that their mass ratio is $q = 1.00 \pm  0.03$, leading to individual masses for B and C of $ 0.241 \pm  0.072\ M_\odot$.  

By interpolating the measured photometry of J1036 with model isochrones of similar age as the UMa MG of 500 Myr we procure masses of the BC pair that are roughly $30\%$ lower than the dynamical estimate (cf. Table~\ref{tab:masses}). The discrepancy is within $1\ \sigma$ of the dynamical mass, but the error is dominated by the distance uncertainty which also affects the luminosity and the theoretical mass. This difference is similar to that found by \citet{Montet} for GJ 3305 B, where the comparison shows a $20\%$ decrease from dynamical to theoretical mass. \citet{Montet} also point out that the $20\%$ discrepancy they obtain for the mass of GJ 3305 B could possibly be explained by an unseen very low-mass companion, a scenario which cannot be ruled out for J1036, albeit unlikely considering that the two equally-mass components B and C would require an additional unseen companion each. 

The discrepancy we find also makes an intriguing comparison to the UMaG member and K-dwarf binary, NO UMa \citep{Schlieder}. This system is in the same moving group as J1036, has the same age, and allows the rare comparison of measured and theoretical masses for coeval low-mass stars.  \citet{Schlieder} measure the mass of the lower-mass, K6.5 type companion, NO UMa B, to be $M_{\rm B} = 0.64 \pm 0.02 M_\odot$. This matches well with the theoretical mass predicted by the BHAC15 500 Myr isochrone of $0.67 \pm 0.01 M_\odot$. In contrast, the dynamical mass estimated here for J1036A, a star of only slightly later-type, appears discrepant from the model predictions. This may be an indicator that the discrepancy between dynamical and theoretical masses becomes more severe when transitioning from late K-dwarfs to early M-dwarfs. Tighter constraints on the components of  J1036 can shed more light on this comparison.

Because the system is fairly close, $\lesssim 20$ pc away, we can assume negligible extinction between us and the system. However, we cannot rule out the possibility of internal extinction within the J1036 system, albeit unlikely as the components B and C are of equal brightness.  In our efforts to produce an SED analysis of the system we conclude that there is a suggestion towards some reddening. This is consistently shown in the $z'$ band, both in our SED analysis as well as comparison to the evolutionary tracks, which may also indicate to some ambiguity with the $z'$ band observations themselves that cannot be ruled out. 

In the near future we expect \emph{Gaia} to provide improved parallactic distance measurement to the system \citep{Gaia Release}, which in turn will put further constrains on the distance and thereby individual masses of its components.  Another possibility is to use long baseline interferometry together with nearby quasars as reference frames to establish a more precise parallactic distance to the system. We also advocate for new resolved observations of J1036, which would benefit to constrain the orbit and relative mass of the system in the coming years, especially in early 2020 when C is predicted to be close to the periastron relative to B.

\begin{acknowledgements} 

MJ gratefully acknowledges funding from the Knut and  Alice  Wallenberg  foundation.
S.D. acknowledges support from the Northern Ireland Department of Education and Learning.

\end{acknowledgements}

\bibliographystyle{aa-note} 
\bibliography{J1036bib}      

\end{document}